\documentclass[sigconf]{acmart}
\usepackage{natbib}
\usepackage{todonotes}
\usepackage{soul}
\usepackage{subcaption} 
\usepackage{balance}

\AtBeginDocument{%
  \providecommand\BibTeX{{%
    \normalfont B\kern-0.5em{\scshape i\kern-0.25em b}\kern-0.8em\TeX}}}

\copyrightyear{2023}
\acmYear{2023}
\setcopyright{acmlicensed}\acmConference[FIRE 2023]{Forum for Information Retrieval Evaluation}{December 15--18, 2023}{Panjim, India}
\acmBooktitle{Forum for Information Retrieval Evaluation (FIRE 2023), December 15--18, 2023, Panjim, India}
\acmPrice{15.00}
\acmDOI{10.1145/3632754.3632760}
\acmISBN{979-8-4007-1632-4/23/12}

\begin{document}

\title{A Comparative Analysis of Retrievability and PageRank Measures} 


\author{Aman Sinha, Priyanshu Raj Mall, Dwaipayan Roy}
\affiliation{%
  \institution{Indian Institute of Science Education and Research, Kolkata, India}
  \city{}
  \country{}
}
\email{as18ms065@iiserkol.ac.in, prm18ms118@iiserkol.ac.in, dwaipayan.roy@iiserkol.ac.in}

\renewcommand{\shortauthors}{Sinha et al.}

\begin{abstract}
The accessibility of documents within a collection holds a pivotal role in Information Retrieval, signifying the ease of locating specific content in a collection of documents.
This accessibility can be achieved via two distinct avenues. The first is through some retrieval model using a keyword or other feature-based search, and the other is where a document can be navigated using links associated with them, if available.
Metrics such as PageRank, Hub, and Authority illuminate the pathways through which documents can be discovered within the network of content while the concept of Retrievability is used to quantify the ease with which a document can be found by a retrieval model.
In this paper, we compare these two perspectives, PageRank and retrievability, as they quantify the importance and discoverability of content in a corpus. 
Through empirical experimentation on benchmark datasets, we demonstrate a subtle similarity between retrievability and PageRank particularly distinguishable for larger datasets.
\end{abstract}




\maketitle

\section{Introduction}\label{sec:intro}

Accessibility of documents within a collection serves as a critical facet of information retrieval, specifying the ease with which documents can be located amidst an extensive corpus.
Essentially, there exist two primary avenues through which this accessibility is assessed.
The first path navigates the terrain of retrieval models, ushering us into the realm of retrievability scores~\cite{azzopardi2008retrievability}. Here, the primary concern is to ascertain whether a particular document can be retrieved within a rank cut off from the vast expanse of a collection by some retrieval model. 
Informally, the retrievability scores quantify how efficiently a document can be retrieved within the collection. It essentially answers the question: how quickly can you pinpoint the proverbial needle in the haystack of documents?
On a parallel course, the accessibility of a document can also be viewed from the point of view of navigability. 
In this context, the focus is not directed at individual documents but rather towards the intricate web of connections between documents and relationships that interlink them.
Navigability, characterized by metrics such as PageRank~\cite{pagerank}, Hub, and Authority~\cite{hits}, illuminates the pathways through which documents can be found.
In this scenario, the focus is not merely on retrieval but on traversing the internal network of documents.
Navigability metrics, such as PageRank, emphasize not just the inherent content of a document, but also its position and importance within the broader context of the document network. 
This metric, distinct from retrievability scores, offers insights into how discoverable a document is through journeys across links and connections.

Very few works have been done in the field to compare retrievability and PageRank.
To the best of our knowledge, the only systematic study was done in~\cite{wilkie2013initial} where only 2K documents are considered for the study in a closed set of webpages from a university website.
Considering both retrievability and PageRank are designed to quantify the discoverability or accessibility (in terms of importance) of contents in a corpus of documents, in this paper, we investigate their alignment through a comparative analysis. 

The rest of the paper is organized as follows.
We present the related work in the next section highlighting the concept of retrievability and PageRank together with some of their applications in the domain of information retrieval before representing the motivation for this work.
We report the empirical results on two benchmark datasets in Section~\ref{sec:eval} accompanied by a comprehensive analysis of the results.
The paper is concluded in Section~\ref{sec:conclusion} mentioning the overall finding and mentioning some future work.

\section{Background and related work}

\subsection{PageRank - a measure of importance}

PageRank is a link analysis algorithm developed by Brin and Page~\cite{pagerank}. Given a set of hyperlinked documents (such as the World Wide Web), the algorithm assigns a numerical weighting to each page of the set. Based on this weight, the relative importance of the pages is measured within the set. Informally, PageRank considers links to be like `votes' by all the other pages on the Web, about how important a page is. A link to a page counts as a vote of support. In addition, it considers that some votes are more important than others. When utilized as a ranking criterion (such as in Google), documents with greater PageRank values are ranked higher in the ranked list.

Formally, the PageRank algorithm is presented in Equation~\ref{eq:pagerank}.
\begin{equation} \label{eq:pagerank}
    PR(A) = (1-d) + d \cdot \sum_{i=1}^{n} \frac{PR(T_i)}{C(T_i)}
\end{equation}

\begin{itemize}
    \item $PR(T_i)$: the self-importance of the webpage $T_i$;
    \item $C(T_i)$: the number of outgoing links from webpage $T_i$;
    \item $\frac{PR(T_i)}{C(T_i)}$: if our page (page A) has a backlink from page $i$, the share of the vote webpage $A$ will get; 
    \item $d$: the damping factor in PageRank helps balance the influence of following links on the current page with the randomness of jumping to other pages, making the PageRank algorithm more realistic and reflective of how web users navigate the internet; traditionally it is set to $0.85$.
\end{itemize}

PageRank does not consider the content or size of a document, the language of the document, or the surrounding text used as the anchor to a link. It only captures the authoritative feature of linked documents which is proven useful for different tasks from text matching~\cite{pang2021pagerank} to word sense dismbiguation~\cite{sheikh2021integrating} although it was first introduced to rank web pages in the Google search engine. 
Further, researchers have used it diverse sub-field of research to improve various downstream tasks.
PageRank has been used as a factor in ranking in~\cite{ricardo2006pagerank}.
It is also employed in~\cite{pang2021pagerank} as a hierarchical noise filtering approach for the long-form text matching problem to filter out noisy information.
The authors plug the PageRank algorithm into the Transformer, to identify and filter both sentence and word-level noisy information in the matching process.

In~\cite{Holzmann2019}, the authors focused on the problem of the deviations in PageRank values caused by restricted crawling.
Some further variation of traditional PageRank is proposed in~\cite{npagerank} replacing the original transition matrix is replaced with one whose entries are based on the number of a node’s N-step neighbours.
PageRank has been utilized in~\cite{mihalcea-tarau-2004-textrank} to extract and score keywords from text documents based on their co-occurrence and position.
It has also been employed for sentiment analysis to extract and rank opinion words and phrases from online reviews in~\cite{kobayashi-etal-2007-extracting}.
\citeauthor{gleich2015pagerank} shows how PageRank can be applied to any graph or network in any domain, such as bibliometrics, social and information network analysis, and link prediction and recommendation.

A comprehensive survey on the applications of PageRank algorithms in various domains can be found in~\cite{chung2014survey,park2019survey}.

\subsection{Retrievability - a measure of accessibility}\label{subsec:ret}
Retrievability, as a metric, gauges the ease with which a document can be retrieved within a specific configuration of an information retrieval (IR) system. The concept of retrievability, formally introduced by Azzopardi and Vinay~[1], is quantified through the retrievability score, denoted as $r(d)$, for a document $d$ within a collection $D$ concerning a particular IR system.
Mathematically, the retrievability score $r(d)$ for a document $d$ ($d \in D$) within the context of an IR system is computed using the formula depicted in Equation~\ref{eq:ret}.
\begin{equation} \label{eq:ret}
    r(d) = \sum_{q \in Q} o_q \cdot f(k_{dq}, c)
\end{equation}

As illustrated in Equation~\ref{eq:ret}, the computation of a document's retrievability relies on an extensive set of queries denoted as $\mathsf{Q}$. This set theoretically encompasses all conceivable queries that could be answered by the collection $D$. Each query $q$ is associated with an opportunity weight $o_q$, which quantifies the likelihood of selecting query $q$ from the query set $\mathsf{Q}$.
The retrieval rank of document $d$ for a particular query $q$ is denoted as $k_{dq}$, and the utility function $f(k_{dq},c)$ serves as an indicator of document $d$'s retrievability within a specified rank cutoff $c$.

The conventional approach for assessing retrievability relies on a cumulative-based approximation, where the utility function $f(k_{dq},c)$ is designed to yield a value of $1$ if document $d$ is retrieved within the top $c$ documents for query $q$, and $0$ otherwise. This utility function offers a straightforward interpretation of the retrievability score for each document. Essentially, it quantifies how frequently the document appears within the top $c$ rankings of various queries. Documents that fall beyond the top $c$ positions are excluded from consideration, replicating a user's behavior when examining only the first $c$ search results. Consequently, a higher retrievability score indicates that the document is retrieved within the top ranks for a larger number of queries.

In order to examine the retrievability bias present in a collection, we can calculate retrievability scores for each document using equation (\ref{eq:ret}). By utilizing the Lorenz Curve, which represents the cumulative score distribution of documents sorted by their retrievability scores in ascending order, we can analyze the degree of inequality or bias within the retrieval system. If retrievability scores are evenly distributed, the Lorenz Curve will be linear. However, a skewed curve indicates a greater level of inequality or bias. To summarize the amount of bias in the Lorenz Curve, the Gini coefficient $G$ is commonly employed~\cite{azzopardi2008retrievability,bashir2009improving,bashir2010improving}, which is computed as shown:
\begin{equation}
    G = \frac{\sum_{i=1}^{N} (2i-N-1)\cdot r(d_i)}{N \sum_{j=1}^{N} r(d_j)}
\end{equation}
Here, $N$ represents total number of documents in the collection.

The Gini coefficient is a measure of inequality within a population~\cite{gini1936measure}. A Gini coefficient of zero denotes perfect equality, indicating that all documents in the collection have an equal retrievability score according to $r(d)$. Conversely, a Gini coefficient of one indicates total inequality, with only one document having $r(d) = |Q|$ while all other documents have $r(d) = 0$. In most cases, retrievability scores exhibit varying degrees of inequality, resulting in a Gini coefficient between zero and one. Consequently, the Gini coefficient provides valuable insights into the level of inequality among documents in terms of their retrievability using a specific retrieval system and configuration. By comparing the Gini coefficients obtained from different retrieval methods, we can analyze the retrievability bias imposed by the underlying retrieval system on the document collection.

Retrievability, and the underlying theory of retrievability, has found applications in various domains. For instance, it has been used in the development of inverted indexes to enhance the efficiency and performance of retrieval systems by capitalizing on terms that contribute to a document's retrievability~\cite{pickens2010reverted}. Additionally, retrievability has been leveraged to investigate bias in search engines and retrieval systems on the web~\cite{azzopardi2009search} and within patent collections~\cite{bashir2010improving}, leading to improvements in system efficiency during pruning processes~\cite{zheng2009document}.

\subsection{Motivation}
Retrievability scores offer insights into the accessibility of documents within a collection, reflecting their ease of retrieval by an information retrieval system. 
On the other hand, PageRank, a fundamental algorithm in web search, assesses the importance and influence of web pages based on their incoming links. 
While both metrics aim to measure the significance of documents, they do so from distinct perspectives. 
Retrievability primarily considers how easily a document can be retrieved, while PageRank evaluates \emph{navigability} of documents in terms of their popularity and how connected they are within a network. 
Our motivation, in this study, is to compare these two metrics to gain insights into the dynamics of information accessibility and navigability, providing a subtle view of document importance. 
This analysis can be useful in various domains, such as information retrieval, search engine optimization, content ranking etc.

A study has been conducted in~\cite{wilkie2013initial} where \citeauthor{wilkie2013initial} compares the correlation between retrievability and navigability measures such as Hub, PageRank and Authority. 
Experiments conducted on three websites with slightly above 2,000 web pages in total reveal a negligible correlation between PageRank and Retrievability with the highest positive correlation reported to be $0.09$.
However, their study was conducted on a tiny set of institution webpages and the results are not reproducible due to the unavailability of the data.
In this paper, we try to perform a similar study on two sizeable and publicly available datasets.

\section{Empirical comparison of Retrievability and PageRank}\label{sec:eval}

\subsection{Datasets and experimental setup}\label{subsec:dataset}
To conduct an empirical investigation comparing retrievability scores and PageRank values, it is essential that the dataset employed possesses a crucial characteristic - the presence of intra-links connecting the documents within the collection. 
This interconnection among documents is a prerequisite for the computation of the PageRank values.
Without such links, the assessment and comparative study of these important metrics becomes unfeasible and impractical.
For our study, we choose datasets that meet this requirement. We employ the English Wikipedia article dump from February 2023\footnote{\url{https://dumps.wikimedia.org/enwiki}}, an extensive dataset famous for its exhaustive coverage as well as intra-linking structure among articles. 
Additionally, we utilize the WT10g collection~\cite{wt10g}, which not only provides textual content but also includes valuable link information for web pages. 
Overall statistics of the datasets are presented in Table~\ref{tab:dataset}.

While performing the retrievability computation, one major component is the employed query set.
For this study, we use the simulation method proposed in~\cite{azzopardi2008retrievability}.
In this procedure, the terms undergo a series of steps that involve analysis and refinement including stemming, and the removal of stopwords. Terms that appear more than five times within the collection are considered single-term queries. 
Further, two-term queries are generated by pairing consecutive terms that each have a collection frequency of at least 20 occurrences. These generated bigrams are then ranked based on their frequency of appearance, with the top two million selected to form the final set of two-term queries.
Note that, the queries are generated separately for each of the collections, and the respective query sets are exclusively used for retrieval on the collection from which they originate. This ensures that the queries remain contextually relevant to their specific collections, maintaining the integrity of the retrieval process.

During retrieval for computing retrievability scores, we employ the Lucene\footnote{\url{https://lucene.apache.org/}} implementation of the BM25 model, with the default parameter settings. 
This choice aligns with the recommendations made by~\citeauthor{azzopardi2008retrievability} in their initial as well as follow-up works~\cite{azzopardi2009search,azzopardi2010relationship,azzopardi2013towards,azzopardi2014page} on retrievability, ensuring consistency with established best practices.
The only parameter of retrievability $c$ (in Equation~\ref{eq:ret}) is set to $100$ while computing the retrievability scores.

In a similar study conducted in~\cite{wilkie2013initial}, a comparison was made between the hub and authority scores as well within a closed set of 2K documents from a university website. 
In contrast, it is worth noting that Wikipedia articles are structured around topics and categories, differing from the general web graph. 
As a result, the application of hub and authority concepts may not be directly applicable in this context. 
Hence, in our current research, we solely focus on comparing PageRank values as a measure of navigability within the Wikipedia dataset.

\begin{table}[]
    \centering
    \caption{Statistics of the datasets utilised for the study.} \label{tab:dataset}
    \begin{tabular}{lrcr}
    \hline
        \textbf{Dataset} & \textbf{\# documents} & \textbf{Collection Type} & \textbf{\# terms} \\ \hline
        \textbf{WT10G} & 1,692,096 & Web & 9,674,707 \\
        \textbf{Wikipedia} & 6,584,626 & Wiki & 18,797,260 \\ \hline
    \end{tabular}
\end{table}

\begin{table}[]
    \centering
    \caption{Gini Coefficient values for the population of Retrievability and PageRank scores computed in the two datasets.}
    \label{tab:gini_ret_pagerank}
    \begin{tabular}{ccc}
    \toprule
         & \multicolumn{2}{c}{\textbf{Gini Coefficient}} \\ \cmidrule{2-3}
         & \textbf{Retrievability} & \textbf{PageRank} \\ \hline
         \textbf{WT10g} & 0.5371 & 0.6618 \\
         \textbf{Wikipedia} & 0.5380 & 0.7050 \\ \bottomrule
    \end{tabular}
\end{table}

\begin{figure*}
  \begin{subfigure}{0.45\textwidth}
    \includegraphics[width=\linewidth]{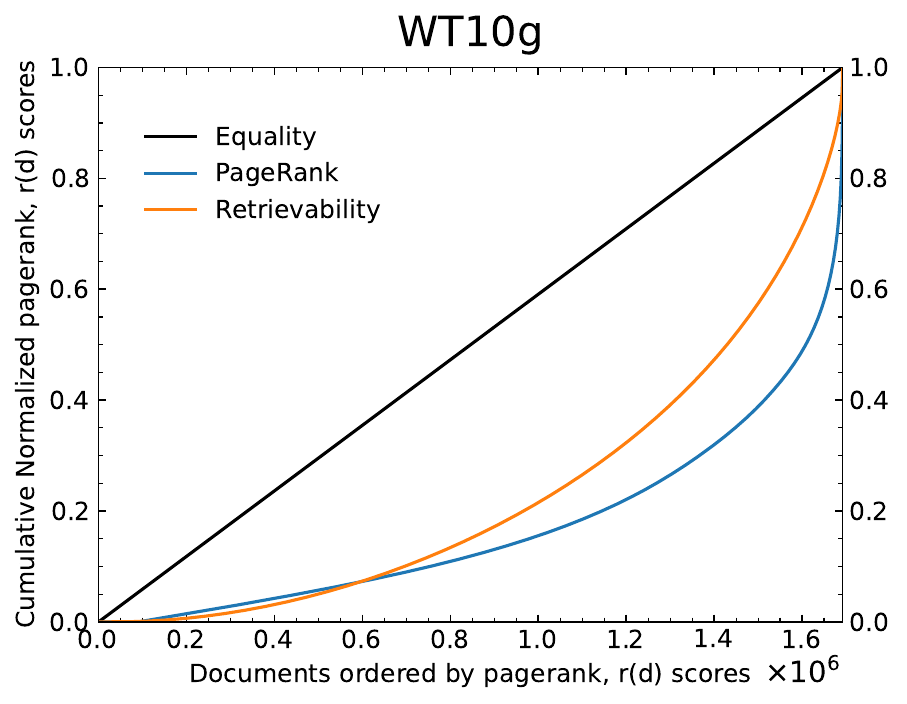}
    \caption{WT10g}
    \label{fig:lc-rb}
  \end{subfigure}
  \begin{subfigure}{0.45\textwidth}
    \includegraphics[width=\linewidth]{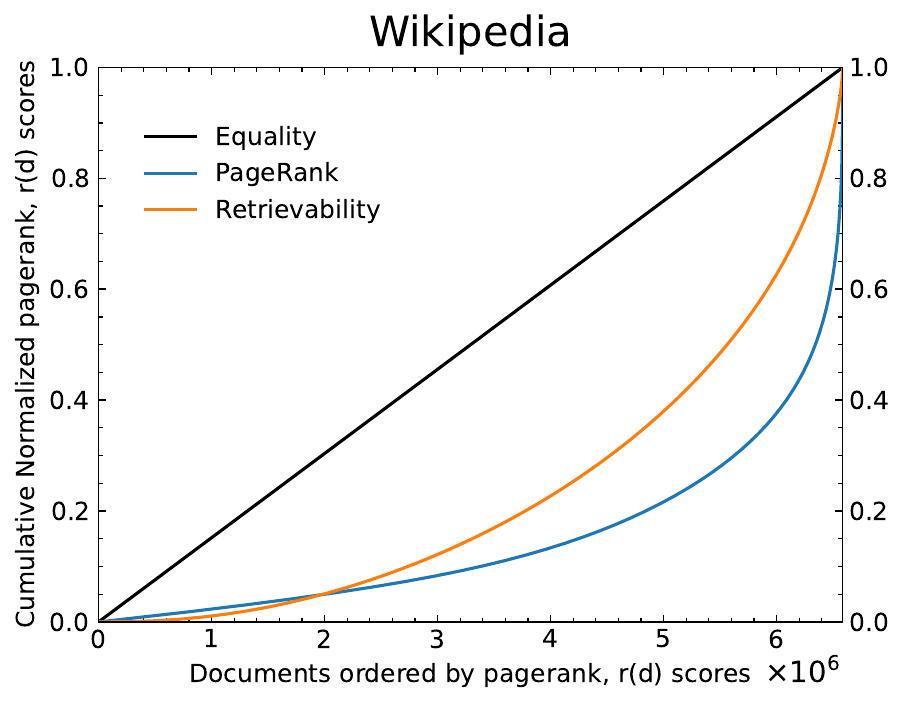}
    \caption{Wikipedia}
    \label{fig:lc-rb}
  \end{subfigure}
  \caption{The Lorenz curve with the distribution of PageRank and Retrievability values on the WT10g collection and the Wikipedia English collection.} \label{fig:lorenz}
\end{figure*}

\subsection{Experimental results and analysis}

In this section, we present the outcomes of our experiments and provide insights drawn from these results. 
Our analysis begins by examining the distribution disparities within the retrievability scores and PageRank values across the datasets we utilized. 
To quantify these disparities, we employ the Gini coefficient, a well-established measure of inequality as discussed in Section~\ref{subsec:ret}. The specific values are presented in Table~\ref{tab:gini_ret_pagerank}.
One notable observation that emerges from this table is the substantial contrast between PageRank values and retrievability scores across datasets. 
This difference is most pronounced in the Wikipedia dataset, where we note a significant $31\%$ difference between PageRank and retrievability values.
The cumulative distributions of both scores are also graphically presented with Lorenz curve in Figure~\ref{fig:lorenz} where the divergence between the PageRank values and the retrievability values becomes specifically apparent in the latter part of the curve. 

In Table~\ref{tab:correl}, we provide correlations between retrievability and PageRank.
To ensure a comprehensive analysis, we employ various rank-based correlation metrics, including Kendall's rank correlation ($\tau$), Spearman's $\rho$, and Ranked Biased Overlap (RBO)~\cite{rbo}. 
Given the inherent differences in the values of retrievability and PageRank due to the way they are computed, we opt for rank-based correlation measures, excluding Pearson's correlation coefficient, which would not be suitable in this context. 

Our analysis reveals a relatively low correlation between the retrievability and PageRank values  indicated by Kendall's $\tau$ of $0.04$ in WT10g collection.
This observation is consistent with the findings from a previous study~\cite{wilkie2013initial}.
Further, Spearman's rank correlation coefficient is noted to be $0.07$ signifying a similar weak positive correlation between these two metrics. 
In contrast, we observe a notable increase in correlation in terms of both Kendall's as well as Spearman's rank correlation coefficient when we extend our analysis to the substantially larger Wikipedia collection.
Specifically, we report correlation coefficients of $0.15$ ($\tau)$ and $0.22$ ($\rho)$ between retrievability scores and PageRank values in the Wikipedia dataset. 
The significant increase in correlation coefficients for
larger dataset suggests that dataset size and content diversity play a substantial role in the relationship between retrievability and PageRank. In other words, retrievability scores and PageRank values tend to exhibit a stronger correlation when working with more extensive and diverse datasets like Wikipedia.
This observation implies that the nature of the documents and their interlinking within the dataset can influence how closely retrievability and PageRank align.

The most interesting insight arises from the value of RBO which exceeds $0.5$ in both the datasets.
This suggests a strong similarity between the rankings of documents when sorted based on their Retrievability and PageRank values.
In essence, while lower Kendall's $\tau$ and Spearman's $\rho$ indicate weak correlations overall, the higher value of RBO reveals a substantial overlap in the top-ranked documents when considering both retrievability and PageRank.
This implies that, although the two metrics may not be highly correlated overall, they tend to agree on at least in terms of the top elements of their respective ranked lists (sorted based on the retrievability and PageRank values).

\begin{table}[]
    \caption{Statistical correlation between Retrievability and PageRank when Retrievability computation is done using the original query generation technique~\cite{azzopardi2008retrievability}.}
    \label{tab:correl}
    \begin{tabular}{l|ccc}
    \toprule
        & \textbf{Kendall's $\tau$} & \textbf{Spearman's $\rho$} & \textbf{RBO} \\ \hline
    \textbf{WT10g}     & 0.0487 & 0.0730 & 0.5173 \\
    \textbf{Wikipedia} & 0.1532 & 0.2247 & 0.5633 \\
    \bottomrule
    \end{tabular}
\end{table}


\section{Conclusion and future work}\label{sec:conclusion}

Given a collection, the accessibility of the documents indicates the ease with which we can find documents which can be dissected based on distinct techniques employed. 
One can use a retrieval model, leading to the computation of retrievability scores, which gauges how readily a document can be retrieved within the collection.
Another avenue involves navigation, where the navigability measures are derived from the interconnections and links between the documents themselves. 
The navigability metrics, such as PageRank, Hub, Authority provide insights into the discoverability via traversing document network and are distinct from the retrievability.
Considering the diverse nature of finding documents based on the two approaches, in this paper, we have done a comparative study of retrievability and PageRank using two web datasets.
Experimentation on WT10g collection reveals an almost negligible correlation between the two metrics in terms of Kendall's and Spearman's correlation coefficient methods.
In contrast, better agreement is observed when Wikipedia, a larger and more extensively linked dataset, is used for the study.
The ranked biased overlap measurements for both datasets show a significant similarity in the ranking of documents sorted based on the respective values.
As part of a future study, a joint measurement of PageRank and Retrievability based on some fusion techniques will be tried.

\bibliographystyle{ACM-Reference-Format}
\balance
\bibliography{refs}

\end{document}